\documentclass{rspublic}

\usepackage{epsfig}

\def\chandra{{\it Chandra\/}}
\def\conx{{\it Constellation-X\/}}
\def\einstein{{\it Einstein\/}}
\def\genx{{\it Generation-X\/}}
\def\hst{{\it {\it HST}\/}}
\def\iso{{\it ISO\/}}
\def\rosat{{\it ROSAT\/}}

\def\sirtf{{\it SIRTF\/}}
\def\uhuru{{\it Uhuru\/}}
\def\xeus{{\it XEUS\/}}
\def\xmm{{\it XMM-Newton\/}}

\def\etal{{\it et~al.\/}}

\def\ltsima{$\; \buildrel < \over \sim \;$}
\def\simlt{\lower.5ex\hbox{\ltsima}}
\def\gtsima{$\; \buildrel > \over \sim \;$}
\def\simgt{\lower.5ex\hbox{\gtsima}}

\begin{document}


\title[The Chandra Deep Field-North Survey]{The Chandra Deep Field-North 
Survey and the Cosmic X-ray Background}

\author[W.N. Brandt, D.M. Alexander, F.E. Bauer, \& A.E. Hornschemeier]
{W. Nielsen Brandt, David M. Alexander, Franz E. Bauer, \& Ann E. Hornschemeier
(for the Chandra Deep Field-North team)}

\affiliation{Department of Astronomy \& Astrophysics, 525 Davey Laboratory, 
The Pennsylvania State University, University Park, PA 16802, USA}

\label{firstpage}

\maketitle

\begin{abstract}{diffuse radiation; surveys; cosmology: observations; 
galaxies: active; X-rays: galaxies; X-rays: general}
\chandra\ has performed a 1.4~Ms survey centred on the Hubble Deep Field-North 
(HDF-N), probing the X-ray Universe 55--550 times deeper than was possible with 
pre-\chandra\ missions. 
We describe the detected point and extended X-ray sources and discuss their 
overall multiwavelength (optical, infrared, submillimeter, and radio)
properties. Special attention is paid to 
the HDF-N X-ray sources, 
luminous infrared starburst galaxies, 
optically faint X-ray sources, and  
high-to-extreme redshift AGN. 
We also describe how stacking analyses have been used to probe the average 
X-ray emission properties of normal and starburst galaxies at cosmologically 
interesting distances.
Finally, we discuss plans to extend the survey and argue that a 5--10~Ms
\chandra\ survey would lay key groundwork for future missions such as 
\xeus\ and \genx. 
\end{abstract}


\section{Introduction}

\subsection{The Cosmic X-ray Background}

The cosmic X-ray background was the first cosmic background radiation 
discovered (Giacconi \etal\ 1962) and has $\approx 10$\% of the energy 
density of the cosmic microwave background (see, e.g., Fabian \& Barcons 1992 
and Hasinger 2000 for reviews). In contrast to the cosmic microwave background,
the cosmic X-ray background is comprised of the integrated 
contributions from a large number of discrete sources. The X-ray background thus 
represents the summed emission from all X-ray sources since the Universe was less than 
a billion years old, and by surveying it we can learn about the nature and evolution 
of X-ray emitting objects over most of the history of the Universe. 

The X-ray background is detected over a broad energy 
band from $\approx$~0.1--200~keV, peaking in 
energy density from 20--40~keV; the logarithmic 
frequency/energy coverage is comparable to that of the far-infrared 
to extreme ultraviolet bandpass. Current deep imaging surveys of the X-ray 
background, however, have been primarily conducted in the narrower
$\approx$~0.1--12~keV band due largely to technological limitations. 

Observations with the new generation of X-ray observatories, \chandra\
(Weisskopf \etal\ 2000) and \xmm\ (Jansen \etal\ 2001), have 
revolutionised studies of the X-ray background and the sources that 
comprise it. Early observations with \chandra\ resolved
most of the 2--8~keV background into point sources (e.g., Brandt \etal\ 2000; 
Mushotzky \etal\ 2000; Giacconi \etal\ 2001); most of the 0.5--2~keV 
background had already been resolved by \rosat\ (e.g., Hasinger \etal\ 1998). 
The accurate positions from the new observatories, particularly \chandra, 
also allow X-ray sources to be matched unambiguously to 
(often faint) multiwavelength counterparts. X-ray surveys 
have finally reached the depths needed to complement the most sensitive 
surveys in the radio, submillimeter, infrared, and optical bands. The
focus has now shifted from simply resolving the X-ray background to 
understanding in detail the sources that comprise it at all X-ray 
fluxes. While it is clear that massive, accreting black holes produce 
the bulk of the X-ray background, the deepest X-ray surveys are now 
allowing the study of other source classes, including starburst and 
normal galaxies. 

In this article, we will review some of the current results from the ongoing 
\chandra\ Deep Field-North (CDF-N) survey. Other ongoing deep X-ray surveys that have 
published results at present include the \chandra\ Deep Field South Survey 
(e.g., Giacconi \etal\ 2002; Rosati \etal\ 2002), the Lockman Hole Survey 
(e.g., Hasinger \etal\ 2001; Lehmann \etal\ 2001), and the SSA13 Survey 
(e.g., Mushotzky \etal\ 2000; Barger \etal\ 2001a). Many other important 
X-ray surveys are being performed over larger solid angles to shallower 
depths. 

Throughout this article we adopt $H_0=65$~km~s$^{-1}$ Mpc$^{-1}$, 
$\Omega_{\rm M}=1/3$, and $\Omega_{\Lambda}=2/3$. 

\subsection{The \chandra\ Deep Field-North Survey}

The CDF-N survey is currently the deepest
X-ray survey ever performed. It is comprised of 1.4~Ms (16.2~days) of 
exposure with the \chandra\ Advanced CCD Imaging Spectrometer (ACIS) covering 
an $\approx 18^\prime\times 22^\prime$ field (see Fig.~1) centred on 
the Hubble Deep Field-North (HDF-N; see Ferguson \etal\ 2000); these
observations are publicly available. This is an 
excellent field for study, since it already has intensive coverage at optical, 
infrared, submillimeter and radio wavelengths (see Ferguson \etal\ 2000
for a review). More than 800 spectroscopic redshifts have been obtained
in the field (e.g., Cohen \etal\ 2000). In the soft (0.5--2~keV) and 
hard (2--8~keV) X-ray bands, the \chandra\ data reach fluxes $\approx 55$ 
and $\approx 550$ times fainter than surveys by previous X-ray 
missions (see Fig.~2). Near the centre of the field, sources with count 
rates as low as $\approx 1$ count per 2~days can be detected. 
Source densities near the soft-band and hard-band flux 
limits are $\approx 7200$~deg$^{-2}$ and $\approx 4200$~deg$^{-2}$ 
(e.g., Brandt \etal\ 2001b; Cowie \etal\ 2002); the fraction of the 
X-ray background resolved in the soft and hard bands is $>90$\% and 
$>80$\%, respectively (the main uncertainty in the resolved fraction
is the absolute normalisation of the background itself). Source 
positions are typically accurate to better than $1^{\prime\prime}$ 
over the entire field, allowing multiwavelength counterparts to be 
identified reliably. 

The CDF-N team includes $\approx 20$ researchers from 
The Pennsylvania State University, 
The University of Hawaii, 
The University of Wisconsin, 
The Massachusetts Institute of Technology, 
The California Institute of Technology, and
Carnegie Mellon University. 
To see all the contributors to the various CDF-N projects, please
refer to the author lists of the cited CDF-N publications. 

\begin{figure}[t!]
\centerline{\psfig{figure=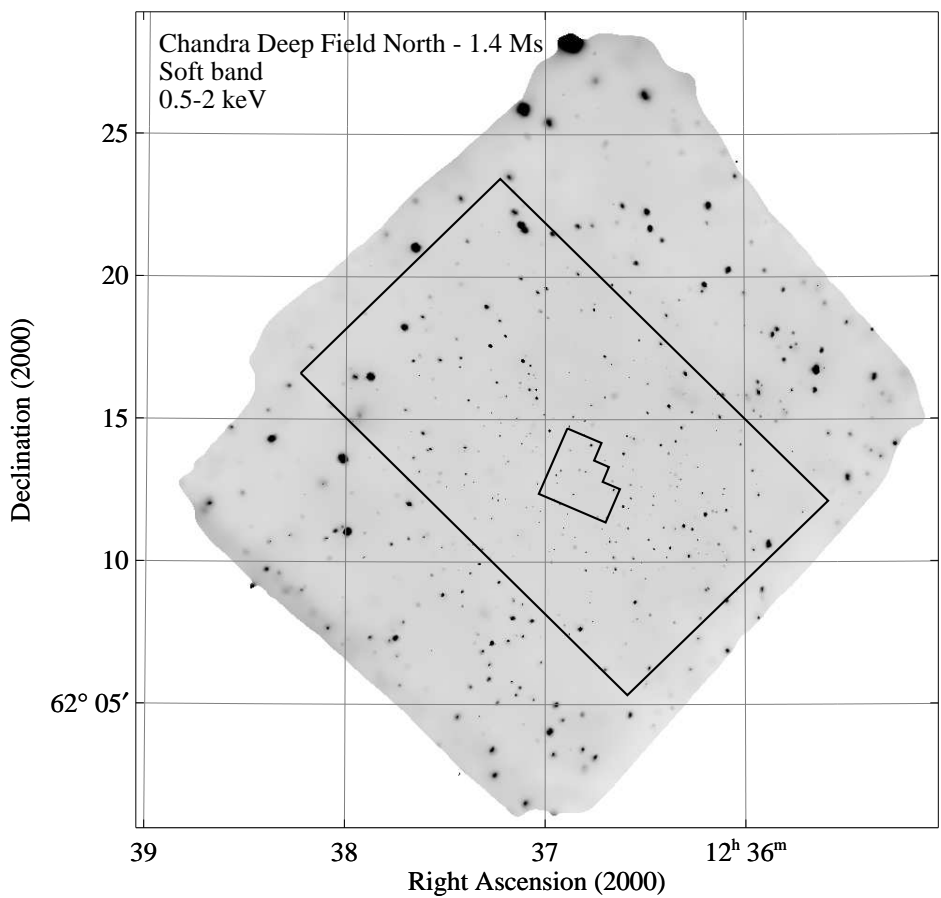,width=5in}}
\caption{Adaptively smoothed and exposure-map corrected image of the 
CDF-N in the soft band. The adaptive smoothing has been performed using 
the code of Ebeling \etal\ (2002) at the $2.5\sigma$ level, and the 
grayscale is linear. 
%
%
Source sizes appear to change across the field due to the spatial 
dependence of the instrumental point spread function. The small polygon 
indicates the HDF-N, and the large rectangle indicates the GOODS area
(see \S3b).}
\end{figure}

\begin{figure}[t!]
\centerline{\psfig{figure=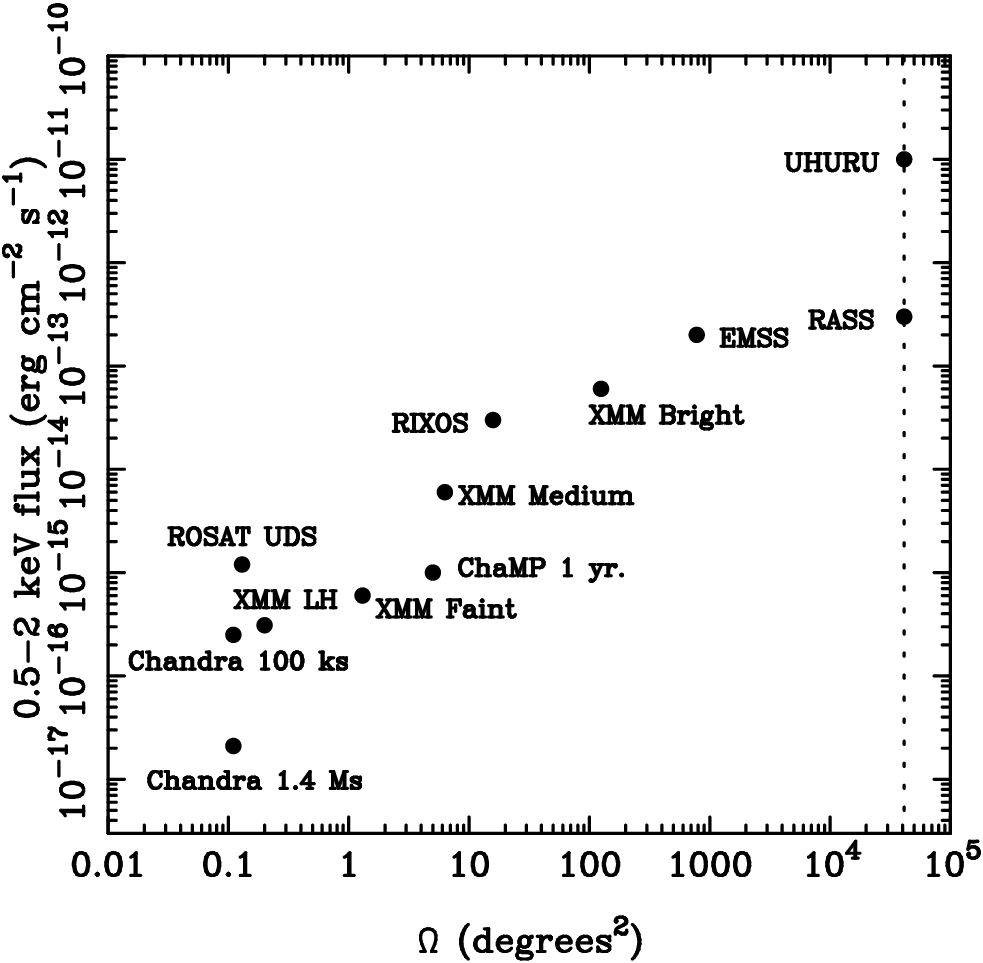,width=4in}}
\caption{A selection of extragalactic X-ray surveys in the 0.5--2~keV 
flux limit versus solid angle, $\Omega$, plane. Shown are 
the \uhuru\ survey,
the \rosat\ All-Sky Survey (RASS), 
the \einstein\ Extended Medium-Sensitivity Survey (EMSS), 
the \rosat\ International X-ray/Optical Survey (RIXOS), 
the \xmm\ Serendipitious Surveys ({\it XMM\/} Bright, {\it XMM\/} Medium, {\it XMM\/} Faint), 
the \chandra\ Multiwavelength Project (ChaMP),
the \rosat\ Ultra Deep Survey (\rosat\ UDS), 
the deep \xmm\ survey of the Lockman Hole ({\it XMM\/} LH), 
\chandra\ 100~ks surveys, and 
\chandra\ 1.4~Ms surveys (i.e., the CDF-N). 
Although each of the surveys shown clearly has a range of flux limits 
across its solid angle, we have generally shown the most sensitive flux limit. 
The vertical dot-dashed line shows the solid angle of the whole sky.
Adapted from Brandt \etal\ (2001b).}
\end{figure}


\section{Some Key Results from the Survey}

\subsection{Detected Point and Extended Sources}

The CDF-N data have been searched intensively for point sources using the 
\chandra\ X-ray Center's {\sc wavdetect} algorithm (Freeman \etal\ 2002). 
At a {\sc wavdetect} false-positive probability threshold of $1\times 10^{-7}$, 
$\approx 430$ independent sources are detected in the 1.4~Ms exposure. 
In addition, a substantial number of sources have been found at lower
significance levels that are spatially correlated with objects found
at other wavelengths (e.g., optically bright galaxies). Many of these
sources are real, but it is more difficult to define a complete X-ray
flux-limited sample of these sources. In the full band (0.5--8~keV), the 
typical source has $\approx 90$ counts which is too small for detailed 
X-ray spectral analysis. Assessment of spectral hardness is possible, 
however, and X-ray spectral analysis can be performed for the brighter 
sources in the field. Catalogs of the detected sources and related analysis 
products have been made publicly available for the 1~Ms exposure (see 
Brandt \etal\ 2001b).\footnote{http://www.astro.psu.edu/users/niel/hdf/hdf-chandra.html}

Six extended X-ray sources have also been detected in the CDF-N 
(Bauer \etal\ 2002). Their X-ray spectral properties, angular 
sizes, and likely luminosities \hbox{($\approx 10^{42}$~erg~s$^{-1}$)} 
are generally consistent with those found for nearby 
groups of galaxies. Two extended X-ray sources of note are (1) a 
group in the HDF-N associated with the $z=1.013$ FR~I 
radio galaxy VLA~J123644.3+621133, and (2) a likely poor-to-moderate 
cluster at $z\simgt 0.7$ that is coincident with an 
overdensity of Very Red Objects (VROs; $I-K\ge4$), optically faint ($I\ge24$) 
radio sources, and optically faint X-ray sources. The surface density of 
extended X-ray sources is $167^{+97}_{-67}$~deg$^{-2}$ at a limiting
soft-band flux of $\approx 3\times 10^{-16}$~erg~cm$^{-2}$~s$^{-1}$. 
No evolution in the X-ray luminosity function of clusters is needed 
to explain this value.

\begin{figure}[t!]
\centerline{\psfig{figure=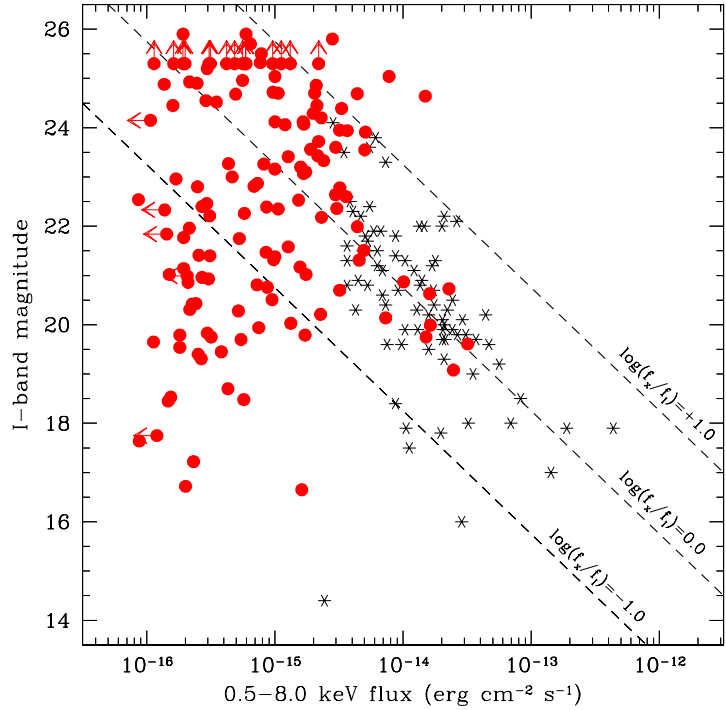,width=4in,angle=0}}
\caption{$I$-band magnitude versus full-band X-ray flux for X-ray 
sources from an $8.4^\prime\times 8.4^\prime$ region of the CDF-N 
field centred on the HDF-N (solid dots; Alexander \etal\ 2001) and 
the \rosat\ ultra deep survey of the Lockman Hole (stars; converted 
to 0.5--8~keV; Lehmann \etal\ 2001). The 
dashed diagonal lines indicate constant X-ray to $I$-band flux 
ratios; luminous AGN typically have $\log(f_{\rm X}/f_{\rm I})$ in 
the range from $-1$ to $+1$. Note the wide spread of $I$-band 
magnitudes for the faintest X-ray sources.}
\end{figure}

\begin{figure}[t!]
\centerline{\psfig{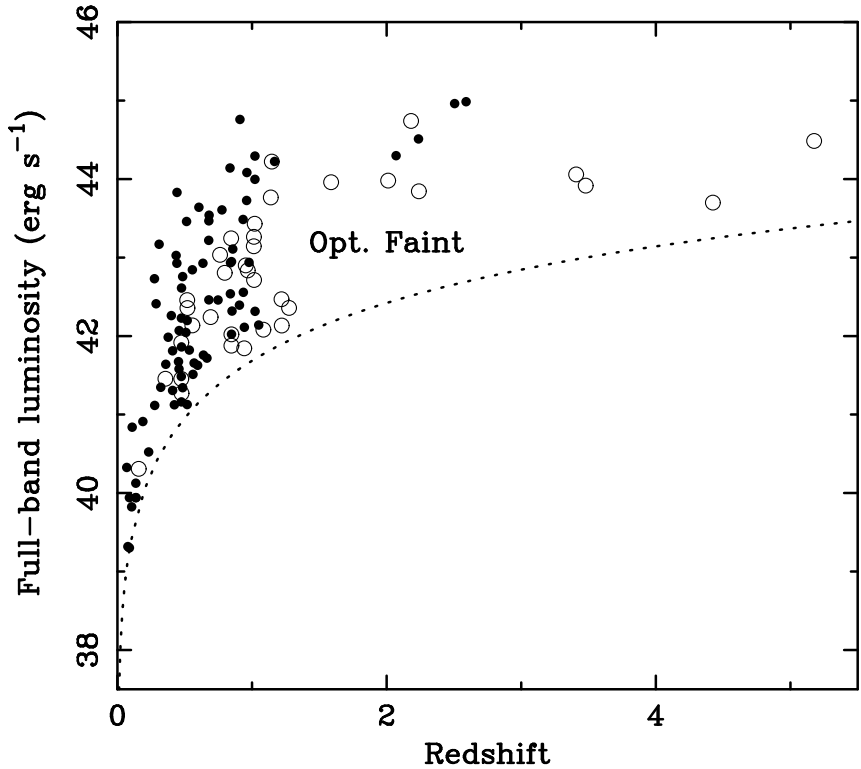}}
\caption{Full-band X-ray luminosity versus redshift for X-ray sources
with optical spectra. Solid dots are sources with $I<21.5$, and
open circles are sources with $I>21.5$. The dotted curve represents
the X-ray detection limit. Note that most of the sources at 
$z\approx$~1--1.3 have $I>21.5$. Similar sources at $z\simgt 1.3$ (in 
the region marked ``Opt. Faint'') will frequently have $I\simgt 24$ 
and thus be too optically faint for spectroscopy (see also Fig.~8b).}
\end{figure}

Fig.~3 shows the $I$-band magnitudes of the X-ray point sources near 
the centre of the CDF-N field. There is a wide spread of $I$-band
magnitudes at faint X-ray fluxes, corresponding to a range of source
types. The faint X-ray sources with \hbox{$I\approx$~15--22} counterparts are 
normal galaxies, starburst galaxies, low-luminosity AGN, and stars. The 
faint X-ray sources with \hbox{$I\simgt 22$} counterparts appear to be 
mostly luminous AGN; these often show evidence for X-ray absorption via 
large hardness ratios. 

Optical spectra are presently available for $\approx 120$ CDF-N 
objects. The majority of the sources with spectroscopic redshifts
lie at $z\simlt 1.3$ (see Fig.~4). This result is at least 
partially due to a selection effect: as implied by Fig.~4, the 
X-ray sources at $z\simgt 1.3$ will frequently be too
optically faint for spectroscopy (with $I\simgt 24$). These
sources are discussed further in \S2d. Even after making
plausible corrections for selection effects, however, it appears
likely that more X-ray power originates from $z\simlt 1.3$ than 
predicted by some X-ray background synthesis models; these models
will require revision. 

\begin{figure}[t!]
\centerline{\psfig{figure=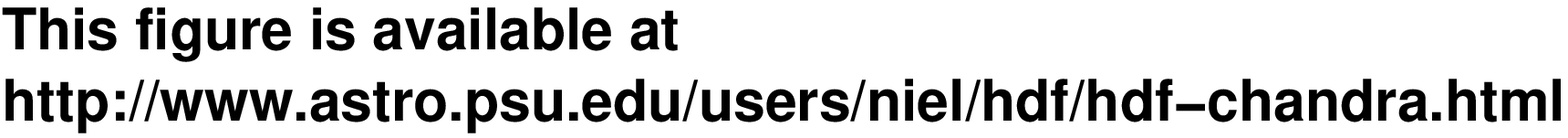,width=4in,angle=0}}
\caption{\chandra\ sources detected in the HDF-N circled on the \hst\ 
optical image. Solid circles indicate sources detected with a 
false-positive probability threshold of $1\times 10^{-7}$, and
broken circles indicate sources found at lower significance levels 
that are spatially correlated with optical galaxies 
(see \S2a). The numbers are source redshifts; redshifts followed
by a ``p'' are photometric rather than spectroscopic. The extended
source in the HDF-N (see \S2a) is located near the $z=1.013$ FR~I 
radio galaxy at the bottom of the image.}
\end{figure}

\subsection{Results for the Hubble Deep Field-North}

Fig.~5 shows the 27 HDF-N X-ray point sources detected thus far
(Hornschemeier \etal\ 2000; Brandt \etal\ 2001a; W.N. Brandt \etal,
in preparation); 16 are found with a false-source probability 
threshold of $1\times 10^{-7}$ while the other 11 are found with lower 
significance but align spatially with 
optical galaxies (see \S2a). As expected from Figs.~3 and 4, 
most sources are at $z<1.5$, and their optical counterparts have a wide
range of brightness. In at least one HDF-N object (CXOHDFN~J123641.7+621131
at $z=0.089$) the X-ray source is offset from the nucleus; the X-ray
emission may arise from a starburst region in the host galaxy as it is
coincident with a bright, blue ``knot.'' 

We have been able to find likely optical or near-infrared counterparts
for all of the X-ray sources in the HDF-N; this contrasts with the case
in the radio where some truly ``blank-field'' sources have been found
(e.g., Richards \etal\ 1999). 
We find a good correspondence between our brighter X-ray sources
and micro-Jy radio (1.4~GHz and 8.5~GHz) sources, but
this trend does not continue for our faintest X-ray sources. For
instance, 10 of the 16 brightest \chandra\ sources have radio
matches, but only 12 of the 27 total \chandra\ sources have radio
matches. The properties of the X-ray/radio sources suggest a broad 
range of emission mechanisms (e.g., Richards \etal\ 1998; Brandt \etal\ 2001a). 
We also find a good correspondence between \chandra\ and \iso\ (6.7~$\mu$m
and 15~$\mu$m; Aussel \etal\ 1999) sources in the HDF-N. Ten of the 
16 brightest \chandra\ sources have \iso\ matches, and 14 of the 27 
total \chandra\ sources have \iso\ matches. This good X-ray/IR correspondence 
bodes well for future IR follow-up of X-ray sources with \sirtf\ (see \S3b). In 
this field with both exceptionally sensitive X-ray and IR coverage, we find a
broad range of X-ray/IR source types including starburst galaxies, 
obscured AGN, and normal elliptical galaxies. In the majority of 
the sources, the IR emission appears to be from dust re-emission of
primary X-ray and ultraviolet radiation. 

\subsection{X-ray Connections with Infrared, Radio and Submillimeter Sources}

\begin{figure}[t!]
\centerline{\psfig{figure=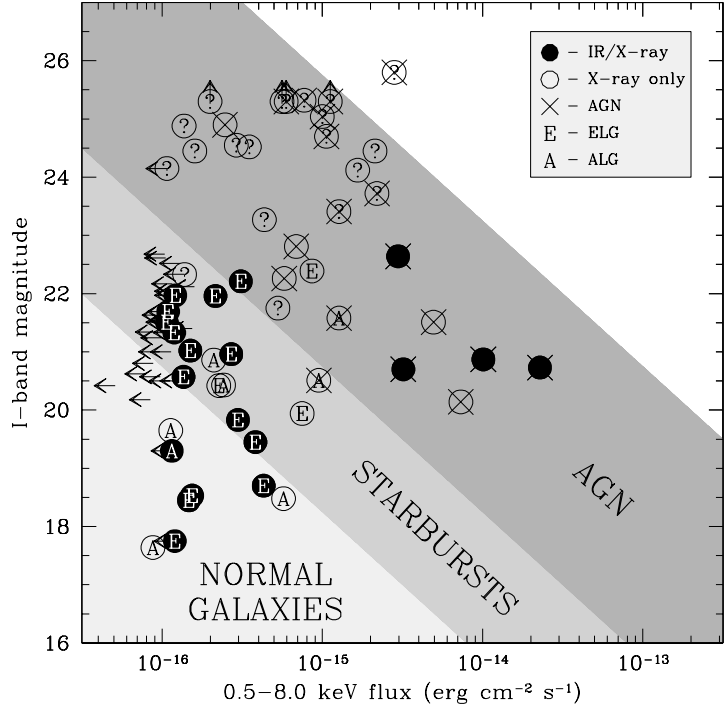,width=4in,angle=0}}
\caption{$I$-band magnitude versus full-band X-ray flux for sources with
both deep \chandra\ and \iso\ coverage. The filled circles are the 
X-ray/IR matched sources, and the open circles are X-ray sources without
IR counterparts. Characters inside the circles indicate different source 
types: ``E'' indicates emission-line galaxies, ``A'' indicates 
absorption-line galaxies, ``?'' indicates sources that do not have 
spectroscopic classifications, and overlaid crosses indicate AGN-dominated 
sources. IR sources without detected  X-ray emission are plotted only as 
upper-limit arrows. The shaded regions delineate the approximate range of 
X-ray-to-optical flux ratio for AGN-dominated sources, starburst galaxies, 
and normal galaxies. Adapted from Alexander \etal\ (2002b).}
\end{figure}

\begin{figure}[t!]
\centerline{\psfig{figure=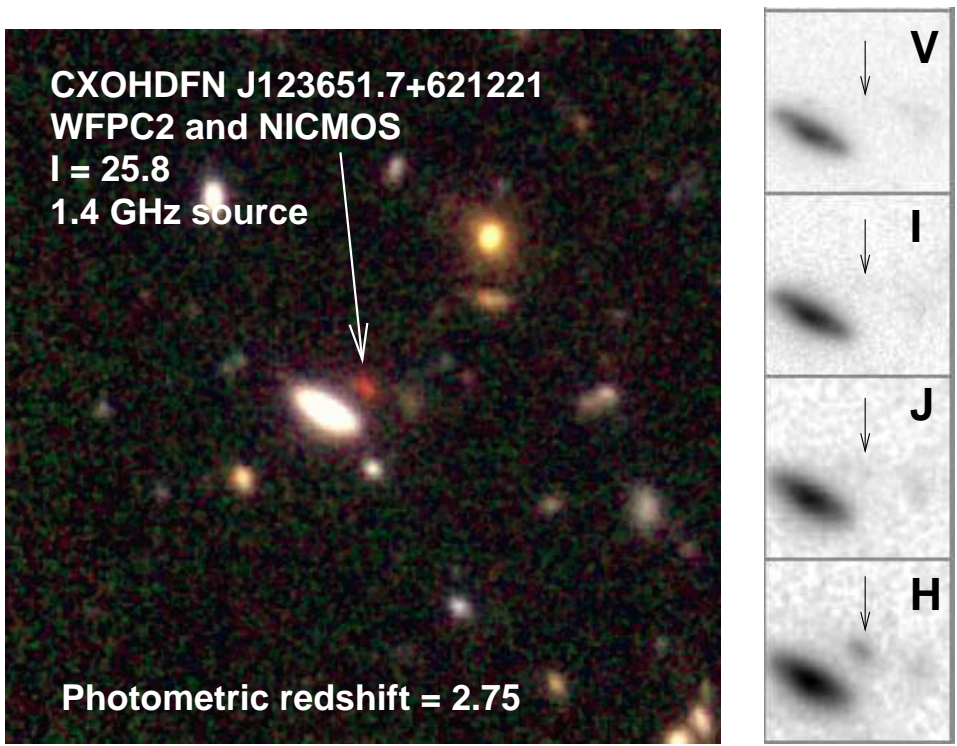,width=5in,angle=0}}
\caption{CXOHDFN~J123651.7+621221, an example of an optically faint 
X-ray source lying in the HDF-N. This $I=25.8$ source has a photometric 
redshift of $z=2.75$ (e.g., Alexander \etal\ 2001) and is remarkably red 
(compare the $V$-band to $H$-band images shown). It is the hardest as well 
as the second X-ray brightest source found in the HDF-N. The X-ray
luminosity ($\approx 3\times 10^{44}$~erg~s$^{-1}$) and hard spectral
shape indicate this source is a luminous, obscured AGN. 
It was previously noted as an optically faint micro-Jy radio source
(e.g., Richards \etal\ 1999) and thought to be a dusty
starburst galaxy (e.g., Muxlow \etal\ 1999). The NICMOS images were 
kindly provided by M. Dickinson (see Dickinson \etal\ 2000).}
\end{figure}

\begin{figure}[t!]
\centerline{\psfig{figure=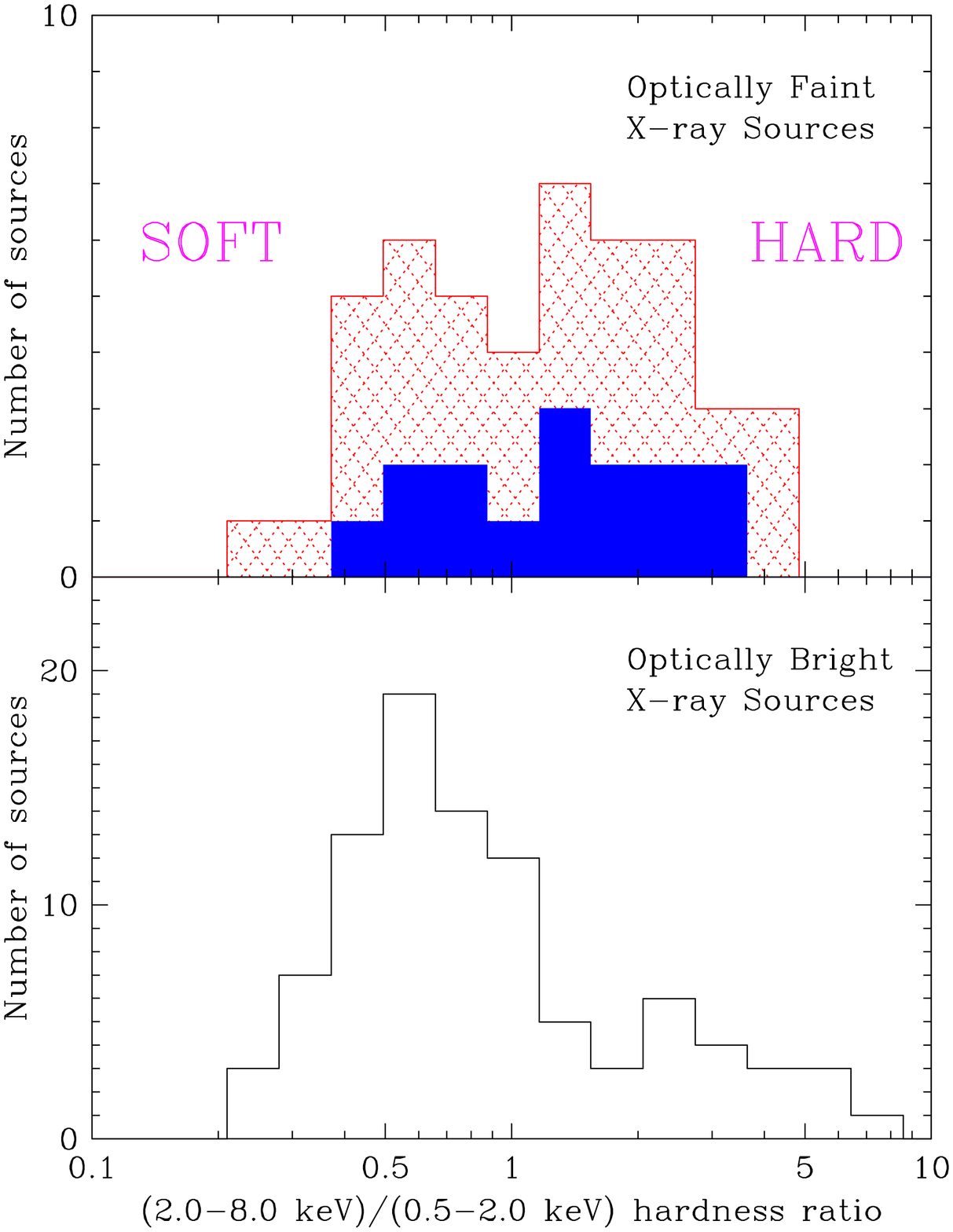,width=2.5in,angle=0}
\psfig{figure=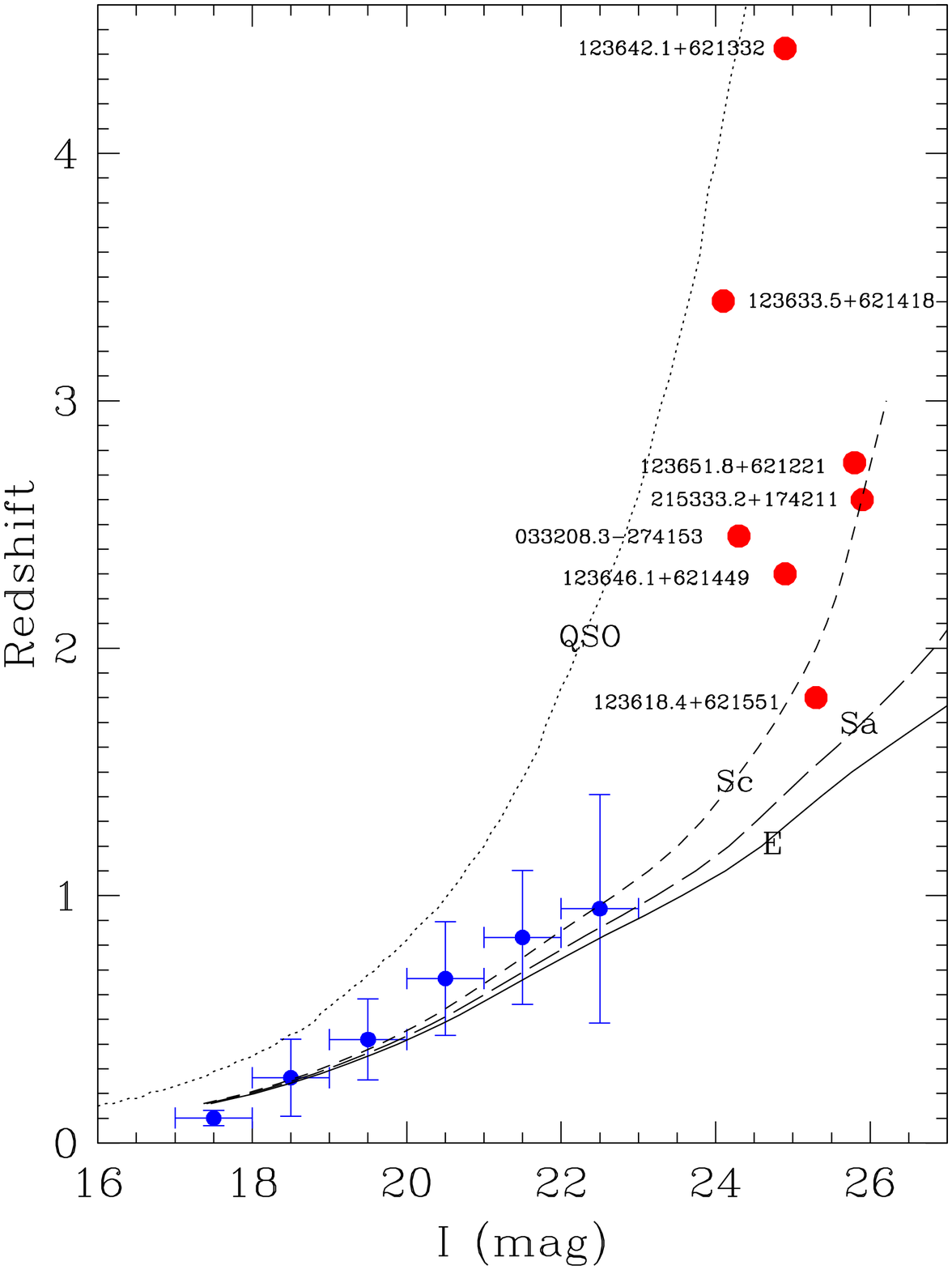,width=2.5in,angle=0}}
\caption{(a) X-ray hardness ratio distributions for optically faint 
($I\ge24$) and optically bright source samples. While both samples
have a wide spread in hardness ratio, the optically faint sources are
harder on average. The solid shading in the upper panel shows the
subset of optically faint sources without $I$-band detections
(corresponding to $I>25.3$). (b) Redshift versus $I$-band magnitude for
X-ray sources. The small solid dots with error bars show the 
average spectroscopic redshifts of the $I<23$ X-ray sources as 
a function of $I$-band magnitude. The large solid dots show individual 
optically faint X-ray sources with spectroscopic, photometric or 
millimetric redshifts. The labelled curves in the diagram show tracks 
for an $M_{\rm I}=-23$ E galaxy, Sa galaxy, Sc galaxy, and QSO. Note 
that the $I<23$ X-ray sources follow the galaxy tracks fairly 
well; their optical emission appears to be dominated by that from the 
host galaxy. If the same holds for the optically faint X-ray sources, 
extrapolation along the curves indicates the majority should lie at 
$z\approx$~1--3. This is indeed consistent with that found for the 
few optically faint sources with redshift determinations. Adapted from 
Alexander \etal\ (2001).}
\end{figure}

More detailed studies of the X-ray/IR sources in the CDF-N have
recently been completed by Alexander \etal\ (2002b) and 
Fadda \etal\ (2002). These use the 21.5~arcmin$^{2}$ region with
uniform \iso\ coverage at 15~$\mu$m rather than just the HDF-N 
(5.3~arcmin$^{2}$). Only $\approx 20$\% of the X-ray/IR sources
are likely AGN. The majority rather appear to be $z\approx$~0.4--1.3
starburst galaxies and $z<0.2$ normal galaxies (see Fig.~6). A 
notable finding is that up to 100\% of the X-ray detected emission-line
galaxies (see Cohen \etal\ 2000 for the optical spectral classification) 
have 15~$\mu$m counterparts (Alexander \etal\ 2002b); the majority of 
these are luminous IR starburst galaxies representative of the population 
making the bulk of the IR background. 
The X-ray/radio matched galaxies trace the same star-forming
population as the X-ray/IR sources with nearly 100\% of the matched galaxies
in common (F.E.~Bauer \etal, in preparation). In addition, the 
X-ray/radio matches appear to trace a population of optically faint 
AGN. 

Barger \etal\ (2001b) have shown that the ensemble of CDF-N X-ray sources 
contribute about 15\% of the extragalactic submillimeter background 
light at 850~$\mu$m, with the strongest submillimeter emission being 
seen from optically faint X-ray sources that are also detected at 
1.4~GHz. At the current CDF-N flux limit, $\approx 20$\% of the 
submillimeter sources have X-ray counterparts. 

\subsection{Optically Faint and High-Redshift X-ray Sources}

As mentioned in \S2a, a significant fraction of the CDF-N X-ray sources
are too optically faint for easy spectroscopic follow-up studies. One
notable example from the HDF-N is shown in Fig.~7. Alexander \etal\ (2001)
present a detailed study of the 47 optically faint X-ray sources
(defined as having $I\ge24$) in an $8.4^\prime\times 8.4^\prime$ 
region centred on the HDF-N. The number of optically faint X-ray 
sources increases at faint X-ray fluxes. However, the fraction of 
optically faint sources within the X-ray source population stays 
roughly constant at $\approx 35$\% for full-band fluxes from 
$10^{-16}$--$10^{-14}$~erg~cm$^{-2}$~s$^{-1}$ due to the emergence 
of a population of optically bright sources (see \S2a and Fig.~3). 
Many of the optically faint X-ray sources have red optical-to-near-IR
colours, and a significant fraction are classified as VROs 
(see also Alexander \etal\ 2002a for a detailed X-ray study of 
CDF-N VROs). They also have harder X-ray spectra on average than 
the sources with $I<24$ (see Fig.~8a). Roughly half of the 
optically faint sources with enough X-ray counts to allow an effective 
search for X-ray variability show it, and the redshifts of the 
majority of the optically faint X-ray sources are estimated to 
be $z\approx$~1--3 (see Fig.~8b). All of these facts support an 
interpretation where most of the optically faint X-ray sources 
are luminous, obscured AGN (e.g., Seyfert~2 galaxies and type~2 
quasars) at intermediate redshifts. 

A minority of the optically faint X-ray sources, however, may
be AGN at high-to-extreme redshifts ($z\approx$~4--10). One notable
example is CXOHDFN~J123642.0+621331 (Brandt \etal\ 2001a; $I=25.3$) 
which lies just outside the HDF-N and is a micro-Jy radio source
at a likely redshift of $z=4.424$ (Waddington \etal\ 1999). With 
an X-ray luminosity of $\approx 5\times 10^{43}$~erg~s$^{-1}$, 
this is by far the lowest luminosity X-ray source known at $z>4$. 
Its detection directly demonstrates that \chandra\ is achieving 
the sensitivity needed to study Seyfert-luminosity AGN at high 
redshift. 

A large population of Seyfert-luminosity AGN at $z\approx$~4--10 
has been postulated by Haiman \& Loeb (1999). These objects would 
represent the first massive black holes to form in the Universe. 
At $z\simgt 6.5$ an AGN should appear not only optically faint but 
optically blank; the Lyman~$\alpha$ forest and Gunn-Peterson
trough will absorb essentially all flux through the $I$ band. 
Thus, an upper limit on the space density of extreme redshift 
AGN can be set simply by counting the number of X-ray sources 
that lack any optical counterpart.\footnote{Of course, this method
requires the plausible assumption that extreme redshift AGN be X-ray 
luminous. The best available data suggest that this should be the
case (e.g., Vignali \etal\ 2001; Brandt \etal\ 2002). Note also 
that at $z\approx 10$ \chandra\ provides rest-frame sensitivity up 
to $\approx 90$~keV; such high-energy X-rays can penetrate a 
substantial amount of obscuration.} Unfortunately, however, confusion 
between truly optically blank sources at extreme 
redshift and very optically faint sources at moderate redshift 
(e.g., objects like that in Fig.~7) occurs without exceptionally
deep optical imaging. At present, the CDF-N data suggest that 
there is $\simlt 1$ AGN detected at $z\simgt 6.5$ per $\approx 12$~arcmin$^{2}$ 
(Alexander \etal. 2001). This limit should soon be tightened 
substantially via the GOODS project (see \S3b). 

\begin{figure}[t!]
\centerline{\psfig{figure=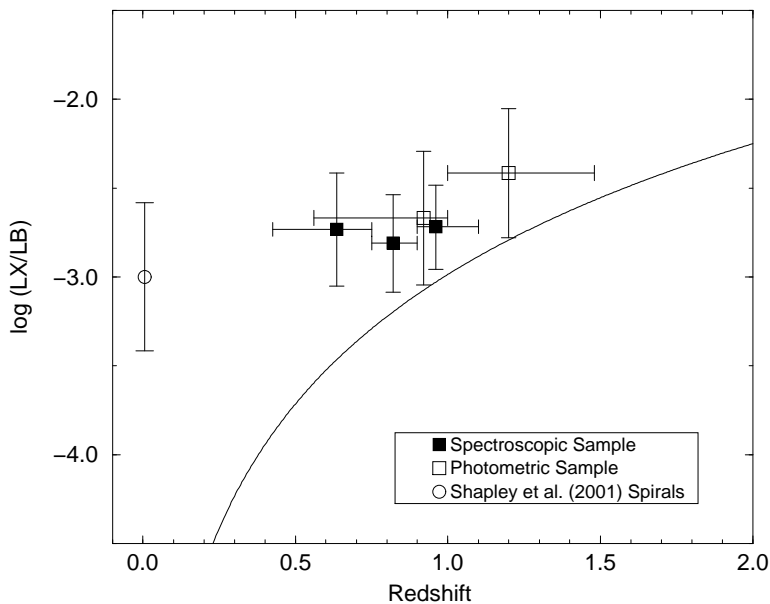,width=4.5in,angle=0}}
\caption{$\log(L_{\rm X}/L_{\rm B})$ as a function of redshift for stacked samples of 
$\approx L_{\rm B}^{\ast}$ spiral galaxies ($L_{\rm X}$ is for 0.5--2~keV); samples 
with both spectroscopically and photometrically derived redshifts are shown. The solid 
curve indicates the $2\sigma$ X-ray sensitivity limit normalized by $L_{\rm B}^{\ast}$. 
The data point at $z\approx 0$ is derived from Shapley \etal\ (2001). Adapted from 
Hornschemeier \etal\ (2002).}
\end{figure}

The highest redshift AGN discovered in CDF-N follow-up studies
has $z=5.18$ (A.J. Barger \etal, in preparation). This is the
highest redshift X-ray selected AGN, and it has relatively 
narrow emission lines. This source is not optically
faint ($I=22.7$) and has an X-ray luminosity of 
$\approx 3\times 10^{44}$~erg~s$^{-1}$; it could have been 
detected by \chandra\ out to $z\approx 10$. 

\subsection{Stacking Studies of Galaxies at Cosmologically Interesting Distances}

\begin{figure}[t!]
\centerline{\psfig{figure=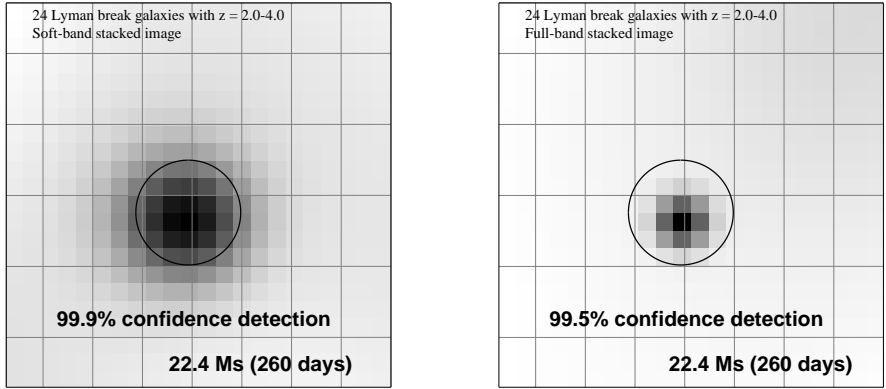,width=5.0in,angle=0}}
\caption{Stacked soft-band and full-band \chandra\ images of 24 HDF-N 
Lyman break galaxies at $z=$~2--4. The black circles are centred on the 
stacking position and have radii of $1.5^{\prime\prime}$. The effective 
exposure time for the average Lyman break galaxy is 22.4~Ms (260~days). 
Both bands give highly significant detections, as assessed with a 
Monte Carlo technique. Adapted from Brandt \etal\ (2001c).}
\end{figure}

At the fainter X-ray fluxes of the CDF-N, a significant number of ``normal'' 
galaxies are detected at $z\simlt 0.3$ where the observed emission appears 
to originate from X-ray binaries, ultraluminous X-ray sources, supernova 
remnants, and perhaps low-luminosity AGN (e.g., Hornschemeier \etal\ 2001;
A.E. Hornschemeier \etal, in preparation). For example, in the HDF-N 
\chandra\ has detected all optically luminous galaxies out to $z=0.15$
(Brandt \etal\ 2001a). 

X-ray studies of normal galaxies at cosmological distances are of 
importance since normal galaxies are expected to be the most numerous 
extragalactic X-ray sources at the faintest X-ray fluxes; in Fig.~5 
there are $\sim 3000$ galaxies but only 27 X-ray sources, so ``there's 
plenty of room at the bottom'' (cf. Feynman 1960). Normal galaxies 
are expected to dominate the number counts at soft-band fluxes of 
$\approx 5\times 10^{-18}$~erg~cm$^{-2}$~s$^{-1}$ (e.g., Ptak \etal\ 2001; 
Hornschemeier \etal\ 2002; Miyaji \& Griffiths 2002), and they should be 
one of the main source types detected by missions such as \xeus\ and \genx. 
Furthermore, the X-ray properties of galaxies might well have changed 
in response to the substantial change in the cosmic star formation 
rate over the history of the Universe; changes in the star formation 
rate should affect the production of X-ray binaries and supernovae 
(e.g., Ghosh \& White 2001).

Most normal galaxies at $z\simgt 0.3$ cannot be individually 
detected in the current CDF-N data, but their average properties 
can be probed using stacking techniques where the X-ray emission 
from many individually undetected galaxies is 
added together. Using stacking techniques, 
Hornschemeier \etal\ (2002) have measured the average X-ray 
luminosities of $\approx L_{\rm B}^{\ast}$ spiral galaxies
out to $z=1.2$, corresponding to a look-back time of 
$\approx 9.0$~Gyr. These measurements allow the first reliable 
predictions of the number of normal galaxies that should be seen 
in deeper X-ray exposures; typical observed soft-band fluxes
are $\approx$~(3--7)$\times 10^{-18}$~erg~cm$^{-2}$~s$^{-1}$.  
There is evidence for a factor of 2--3 increase in X-ray luminosity 
(per unit $B$-band luminosity) with redshift (see Fig.~9), although 
this is weaker than some theoretical models have predicted. 

At higher redshifts ($z=$~2--4), X-ray stacking analyses 
have allowed an average detection of the Lyman break galaxies in 
the HDF-N (see Fig.~10; Brandt \etal\ 2001c). In the rest-frame 
ultraviolet and optical bands these galaxies share many of the 
properties of local starburst galaxies and have estimated star 
formation rates of $\approx$~20--50~M$_\odot$~yr$^{-1}$. 
We find their average X-ray luminosity 
($\approx 3\times 10^{41}$~erg~s$^{-1}$ in the rest-frame
2--8~keV band) to be similar to those of the most luminous local 
starburst galaxies (these have star formation rates comparable to those
estimated for the Lyman break galaxies). For Lyman break galaxies, 
the observed ratio of X-ray to $B$-band luminosity is somewhat,
but not greatly, higher than that seen from local starburst galaxies. 


\section{The Future}

\subsection{Additional X-ray Coverage}

\begin{figure}[t!]
\centerline{\psfig{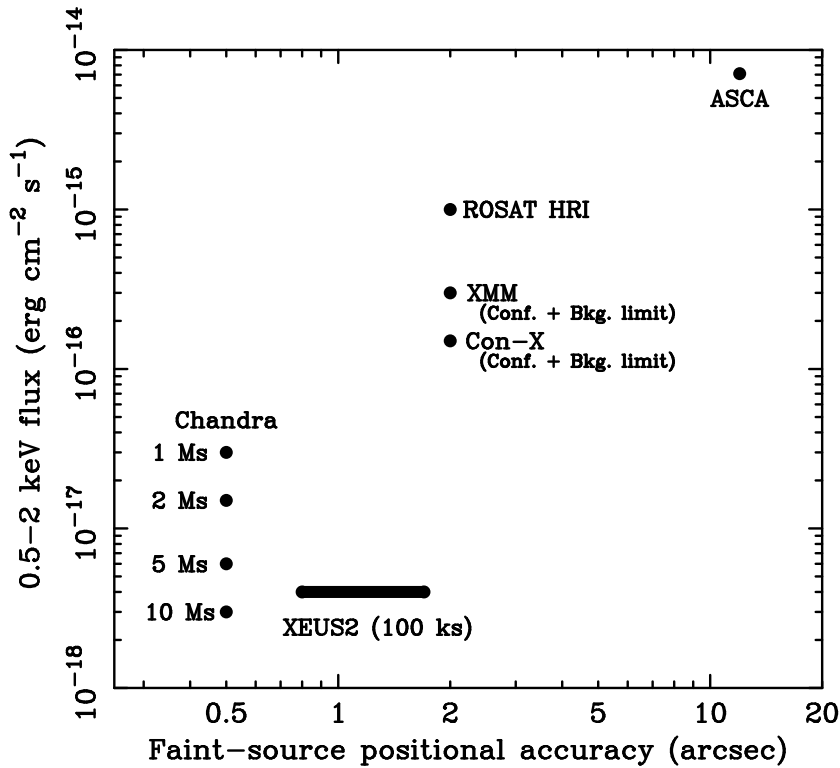}}
\caption{Flux limit versus faint-source positional accuracy for some
past, present, and future X-ray missions. Note that the locations in 
the diagram for future missions should be taken as approximate, and
that \conx\ is focused on high-throughput spectroscopy rather than
deep surveys. Both \xmm\ and \conx\ are background limited and suffer
from source confusion at approximately the positions shown. With 
sufficient exposure, \chandra\ can achieve sensitivities 
comparable to those discussed for future missions 
such as \xeus. Furthermore, \chandra\ positions are likely to be 
the best available for $\simgt$~15--20~years.}
\end{figure}

The current CDF-N survey is far from the limit of \chandra's capability. 
The full-band detector background is so low that, with appropriate grade 
screening, \chandra\ will not fully enter the background-limited regime 
near the aim point for exposure times of $\simlt~5$~Ms. In the soft band
the situation is even better due to the lower background, and \chandra\
should remain photon limited to $\approx 10$~Ms. At present, \chandra\ 
observation time is allocated to extend the survey to 2~Ms, and 
additional observations will be proposed. 

Fig.~11 shows that with sufficient exposure (5--10~Ms), \chandra\ surveys 
can reach depths comparable to those discussed for missions such as \xeus. 
They would thereby bolster the science cases for missions such as \xeus\ 
and \genx, providing key information on the existence and nature of the 
sources to be targeted by these missions. Important targets would 
include the first massive black holes at high redshift as well as normal
and starburst galaxies at intermediate redshift. Detailed spectral, 
temporal, and spatial constraints would be obtained for all the 
sources currently detected. Furthermore, Fig.~11 shows that \chandra\ 
positions are likely to be the best available for $\simgt$~15--20~years; 
these will be essential for the reliable identification of optically 
faint sources at high redshift. 

\subsection{Observations at Other Wavelengths}

Multiwavelength follow-up studies of the CDF-N sources continue, and 
a catalog presenting the current optical photometric and spectroscopic
results will be completed shortly (A.J. Barger \etal, in preparation). 
The Great Observatories Origins Deep Survey (GOODS) project 
will soon obtain deep, public \hst\ Advanced Camera
for Surveys and \sirtf\ coverage over the deepest $\approx 1/3$ of the
CDF-N (see Fig.~1).\footnote{http://www.stsci.edu/science/goods/}
These projects, along with others already completed
and in progress, will provide a panchromatic data set with the sensitivity 
and angular resolution needed to complement fully the CDF-N. 


\begin{acknowledgements}

We thank all the members of the CDF-N team. 
This work would not have been possible without the efforts 
of the entire \chandra\ and ACIS teams. 
We gratefully acknowledge the financial support of
NASA grant NAS~8-38252 (Gordon P. Garmire, PI),
NSF CAREER award AST-9983783 (WNB, DMA, FEB), and  
NASA GSRP grant NGT~5-50247 and the Pennsylvania Space Grant Consortium (AEH). 

\end{acknowledgements}



\label{lastpage}

\end{document}